\documentclass[titlepage,12pt]{article}
\usepackage{amssymb,psfig,epsfig,pslatex}
\usepackage{cite}
\usepackage{epsfig}   %
\usepackage{float}    %
\usepackage{wrapfig}  %
\restylefloat{figure} %

\makeatletter
\def\@sect#1#2#3#4#5#6[#7]#8{\ifnum #2>\c@secnumdepth
  \def\@svsec{}\else
  \refstepcounter{#1}\edef\@svsec{\csname the#1\endcsname.\hskip0.5em}\fi
  \@tempskipa #5\relax
  \ifdim \@tempskipa>\z@
    \begingroup
      #6\relax
      \@hangfrom{\hskip #3\relax\@svsec}{\interlinepenalty \@M #8\par}%
    \endgroup
    \csname #1mark\endcsname{#7}\addcontentsline
      {toc}{#1}{\ifnum #2>\c@secnumdepth \else
        \protect\numberline{\csname the#1\endcsname}\fi #7}%
  \else
    \def\@svsechd{#6\hskip #3\@svsec #8\csname #1mark\endcsname
      {#7}\addcontentsline{toc}{#1}{\ifnum #2>\c@secnumdepth \else
        \protect\numberline{\csname the#1\endcsname}\fi #7}}%
  \fi \@xsect{#5}}
\@addtoreset{equation}{section}
\makeatother
\renewcommand\thesection{\Roman{section}}
\renewcommand\theequation{\ifnum \value{section}>0
 \thesection.\arabic{equation}%
\else
\arabic{equation}%
\fi}
\def\shat{\hat{s}}

\def\sp{{\mathbf S}_t}
\newcommand{\sm}{{\bf S}_{\bar t}}
\newcommand{\kh}{{{\bf \hat k}}}
\newcommand{\ph}{{\bf \hat p}}
\newcommand{\dhh}{{\bf \hat d}}

\newcommand{\be}{\begin{equation}}
\newcommand{\ee}{\end{equation}}

\newcommand{\bea}{\begin{eqnarray}}
\newcommand{\eea}{\end{eqnarray}}
\newcommand{\bfig}{\begin{figure}}
\newcommand{\efig}{\end{figure}}
\newcommand{\bc}{\begin{center}}
\newcommand{\ec}{\end{center}}

\renewcommand{\thefootnote}{\small\fnsymbol{footnote}}

\begin{document}
\begin{titlepage}
  \begin{flushright}
   {\tt hep-ph/0508091}\\
    PITHA 05/03  \\
    SDU-HEP200502
  \end{flushright}        
\vspace{0.01cm}
\begin{center}
{\LARGE {\bf Mixed QCD and weak corrections
to top quark pair production  at hadron colliders}}  \\
\vspace{2cm}
{\large{\bf W. Bernreuther\,$^{a,}$\footnote{Email:
{\tt breuther@physik.rwth-aachen.de}},
M. F\"ucker \,$^{a,}$\footnote{Email:
{\tt fuecker@physik.rwth-aachen.de}},
Z. G. Si\,$^{b,}$\footnote{Email: {\tt zgsi@sdu.edu.cn}} 
}}
\par\vspace{1cm}
$^a$Institut f\"ur Theoretische Physik, RWTH Aachen, 52056 Aachen, Germany\\
$^b$Department of Physics, Shandong University, Jinan, Shandong
250100, China\\
\par\vspace{1cm}
{\bf Abstract}\\
\parbox[t]{\textwidth}
{The order $\alpha_s^2 \alpha$ mixed QCD and weak corrections to top
quark pair production by  quark antiquark annihilation are computed, keeping the
full dependence on the $t$ and $\bar t$ spins. We determine the
contributions to the  cross section and to single and double top spin asymmetries
at the parton level. These results are necessary ingredients for
precise standard model predictions of top quark observables, in
particular of
top spin-induced parity-violating angular
correlations and asymmetries at hadron colliders. }
\end{center}
\vspace*{2cm}

PACS number(s): 12.15.Lk, 12.38.Bx, 13.88.+e, 14.65.Ha\\
Keywords: hadron collider physics, top quarks, QCD and electroweak 
corrections, parity violation, spin effects
\end{titlepage}
%
%
\setcounter{footnote}{0}
\renewcommand{\thefootnote}{\arabic{footnote}}
\setcounter{page}{1}

One promising tool for investigating the so-far
relatively unexplored dynamics of top quark
production and decay, once high 
statistics samples of $t$ and/or $\bar
t$ quarks are available, are observables associated with the spins of
these quarks. As far as   QCD-induced $t\bar t$
production and decay  at hadron colliders is concerned, theoretical predictions for
differential distributions including the full dependence
 on the $t, \bar t$ spins are available at NLO 
in the QCD coupling \cite{Bernreuther:2001rq,Bernreuther:2004jv}.

For full exploration of sizeable, respectively 
large $t \bar t$ data samples that are expected at the Tevatron and
at the LHC the standard model (SM) predictions should be
as precise as possible. Specifically weak interaction contributions to
$t \bar t$ production should be taken into account. Although they are
nominally subdominant with respect to the QCD contributions they can
become important at large $t \bar t$ invariant mass due to large
Sudakov logarithms (for reviews and references, see e.g. 
\cite{Melles:2001ye,Denner:2001mn}).

SM weak interaction effects in hadronic production of heavy quark pairs
were considered previously.  
The parity-even and parity-odd order $\alpha_s^2
\alpha$ vertex corrections\footnote{Here $\alpha_s$ and $\alpha$
  denote  the strong and electromagnetic
couplings, and the weak  coupling is $\alpha_W =\alpha/sin^2\theta_W$.}  were determined 
in \cite{Beenakker:1993yr} and in
\cite{Kao:1999kj}, respectively (see also \cite{Kao:1997bs}). 
In ref. \cite{Kao:1999kj}
also parity-violating non-SM effects were analysed.
The box contributions to $q {\bar q} \to t \bar t$ were not taken into
account in these papers. In \cite{Maina:2003is} the weak contributions to
the  hadronic  $b \bar b$ cross section, including these box
contributions,  were computed. 

In this Letter we report on the calculation of 
the mixed QCD and weak radiative corrections of order $\alpha_s^2
\alpha$ to the  (differential) cross section
of  $t\bar t$ production by quark-antiquark annihilation, keeping the full
information on the spin state of the $t\bar{t}$ system. 
These results are necessary ingredients for definite SM predictions, 
in particular of parity-violating observables associated with the
spin of the (anti)top quark. 

In the following we first give some details of our calculation.
Then we present numerical results for the cross section and for several
single spin and spin-spin correlation observables. 

Top quark pair production both at the Tevatron and at the LHC is
dominated by the QCD contributions to $q \bar q \to t \bar t$ and
$g g \to  t \bar t$, which are known to order $\alpha_s^3$. 
Due to color conservation there are no $\alpha_s \alpha$ Born level
contributions to these processes. The leading contributions involving
electroweak interactions are the order  $\alpha^2$  Born terms  for
$q \bar q \to t \bar t$ and the
mixed contributions of order  $\alpha_s^2 \alpha$.
For the quark-antiquark annihilation processes, which we analyze in
the following, this amounts to studying the reactions
\begin{equation}
q(p_1)+\bar{q}(p_2)\rightarrow t(k_1, s_t)+\bar{t}(k_2, s_{\bar t})\, ,
\label{qqs12}
\end{equation}
\begin{equation}
q(p_1) + {\bar q}(p_2) \rightarrow t(k_1, s_t)+\bar{t}(k_2, s_{\bar t}) + g(k_3),
\label{eq:qqgluon}
\end{equation}
Here $p_1$, $p_2$, $k_1$, $k_2$, and $k_3$ denote the parton momenta.
The vectors $s_t$,  $s_{\bar t}$,
with  $s^2_t = s^2_{\bar t} = -1$ and $ k_1\cdot s_t = 
k_2\cdot  s_{\bar t} = 0$
describe the spin of the top and antitop quarks.
All quarks but the top quark are taken to be massless. 
 
The respective contributions to the differential cross section of
(\ref{qqs12}) are of the form
\be
\alpha^2|{\cal M}_2(p,k,s_t, s_{\bar t})|^2 + 
\alpha_s^2 \alpha \; \delta {\cal
  M}_2(p,k,s_t, s_{\bar t}) \, ,
\label{convir}
\ee
where ${\cal M}_2$ corresponds to the $\gamma$ and $Z$
exchange diagrams. As we are interested in this Letter in particular  in
parity-violating
effects, we take into account only the mixed QCD and weak contributions to  $\delta
{\cal M}_2$ and to (\ref{eq:qqgluon})  
in the following. The photonic  contributions 
form a gauge invariant set and can be straightforwardly
obtained separately. 
The contributions to $\delta {\cal M}_2$ are 
the order $\alpha_s^2$ two-gluon box diagrams
interfering with the Born $Z$-exchange diagram, 
and the $Z$ gluon ($g$) box diagrams and the diagrams with
the weak corrections to the $q {\bar q} g$ and $g t \bar t$ vertices
interfering with  the Born gluon exchange diagram. The ultraviolet
divergences  in the vertex corrections are removed using the
on-shell scheme for defining the wave function renormalizations
of the quarks and the top quark mass $m_t$. \\
The respective contributions to the differential cross section of
(\ref{eq:qqgluon}) are of the form $\alpha_s^2\alpha \,\delta {\cal
  M}_3(p,k,s_t, s_{\bar t})$  and result from the interference of the order 
$g_s^3$ with the order $g_s e^2$ gluon bremsstrahlung diagrams.

The box diagram contributions to
(\ref{convir})  contain infrared divergences due to virtual soft
gluons. They are canceled against terms from soft gluon
bremsstrahlung. As as 
consequence of color conservation  both the sum of the box diagram
contributions to $\delta {\cal M}_2$ and $\delta {\cal M}_3$ are free
of collinear divergences.  \\
We have  extracted the IR divergences, using dimensional
regularization, with two different methods: a phase space slicing
procedure (as in \cite{Bernreuther:2004jv}) and, alternatively as a check, we
have  constructed subtraction  terms that render the three particle
phase space integral over the subtracted term  $[\delta {\cal M}_3]_{subtr}$ finite. 
When calculating observables, in particular those given below, both
methods led to results which numerically agree to high precision.  

We have determined  (\ref{convir}) and $\delta {\cal M}_3$,
respectively their infrared-finite counterparts, analytically for
arbritrary $t$ and $\bar t$ spin states. From these expressions one
may extract the respective production spin density matrices. These matrices, combined with 
the decay density matrices
describing
semi- and non-leptonic $t$ and $\bar t$ decay
\cite{Brandenburg:2002xr}
then yield, in the $t \bar t$ leading pole or narrow width approximation,
 standard model predictions for  distributions 
of the reactions $q {\bar q} \to t {\bar t} \to b {\bar b} + 4 f \, (+g)$ 
($f=q,\ell, \nu_\ell$) with the $t$
and $\bar t$ spin degrees of freedom fully taken into account. 

The expressions for (\ref{convir}) and $\delta {\cal M}_3$ are rather
lengthy when the full dependence
on the $t$ and $\bar t$ spins is kept, and we do not reproduce them
here. 
We represent  these
contributions to the partonic cross sections  and to several single and double spin
asymmetries, which we believe are of interest to phenomenology, in
terms of dimensionless scaling functions depending on the kinematic variable 
$\eta = \frac{\shat}{4m^2_t} -1$, where ${\shat}$ is the $q \bar q$
c.m. energy squared. The inclusive, spin-summed  $ q \bar q$ cross sections 
for  (\ref{qqs12}), (\ref{eq:qqgluon}) may be written,
to NLO in the SM couplings, in the form 
\begin{equation}
{\sigma}_{q \bar q} \; = \; {\sigma}_{q \bar q}^{(0)QCD} + 
\delta {\sigma}_{q \bar q}^{QCD} + \delta {\sigma}_{q \bar q}^{W} \; ,
\label{mixed}
\end{equation}
where the first and  second term are  the LO (order $\alpha_s^2)$ and
NLO (order $\alpha_s^3)$ QCD contributions 
\cite{Nason:1987xz,Beenakker:1990ma},  \cite{Bernreuther:2000yn}, and the third term is
generated by the electroweak contributions 
  (\ref{convir}) and
$\delta {\cal M}_3$ described above. We decompose this term
as follows:
\begin{equation}
\delta {\sigma}_{q \bar q}^W(\shat,m^2_t) \; = \; \frac{4\pi \alpha}{m^2_t}[
\alpha \, f^{(0)}_{q \bar q}(\eta) + \alpha_s^2 \, f^{(1)}_{q \bar q}(\eta)]
\, .
\label{eq:xsection}
\end{equation}
We have  numerically evaluated the scaling functions  $f^{(i)}(\eta)$ 
 -- and those given below -- and parameterized them in terms of fits
 which allow for a quick  use in applications. 
In the following we use $m_t = 178$ GeV, $m_Z=91.188$ GeV, and 
$sin^2\theta_W =0.231$. In  Figure 1 and Figs. 3 - 9 below we use
 $m_H =114$ GeV for the mass of the standard model Higgs boson. 
The dependence  on  the Higgs boson mass 
is shown in Fig. 2 in the case of  $f^{(1)}_{d \bar d}$  for two values
of $m_H$  \cite{unknown:2004qh}.

\begin{figure}[H]
\begin{center}
\epsfig{file=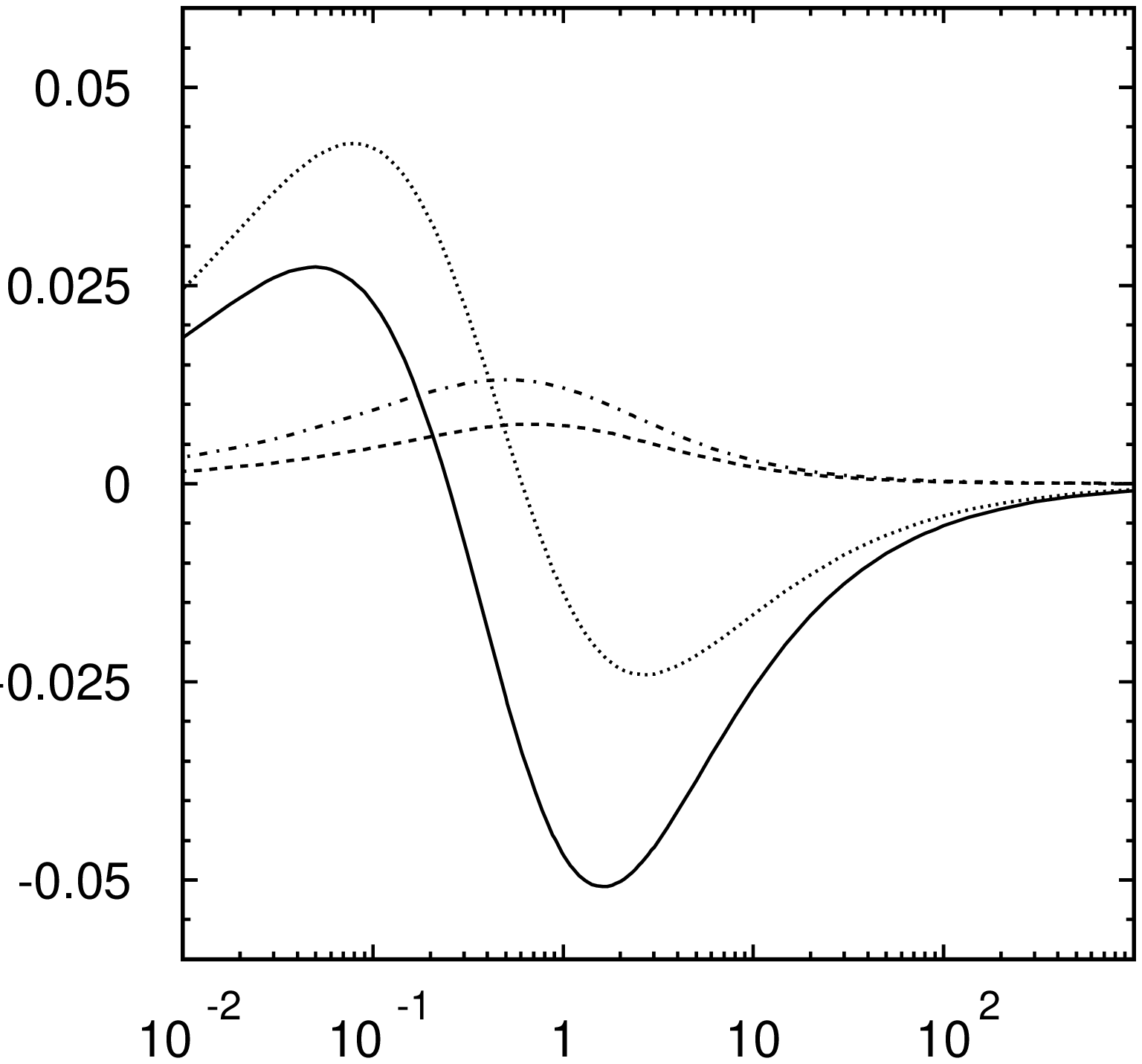, width=12cm ,height=8cm}
\end{center}
\caption{Dimensionless scaling functions $f^{(0)}_{q \bar q}(\eta)$
(dashed), $f^{(1)}_{q \bar q}(\eta)$ (solid)  that determine
the parton cross section  (\ref{eq:xsection}) for
$q=d$ type quarks. The dash-dotted and dotted lines correspond to
the respective functions for $q=u$ type quarks. The Higgs
boson mass is put to 114 GeV.}\label{fig:sigqq}
\end{figure}

In Fig. \ref{fig:sigqq} the  functions $f^{(i)}_{q \bar q}$ are
displayed 
as functions of $\eta$ for 
annihilation of initial massless partons $q \bar q$ of the first and
second generation with weak isospin $\pm 1/2$. As expected the 
$\alpha_s^2 \alpha$ corrections are significantly larger than the the lowest
order photon and $Z$ boson exchange contributions. 
The correction (\ref{eq:xsection}) to the
$q \bar q$ cross section  has recently been computed also 
by \cite{KSU}. We have compared our results and find excellent
numerical agreement. 

\begin{figure}[H]
\begin{center}
\hspace*{-1.5cm} \epsfig{file=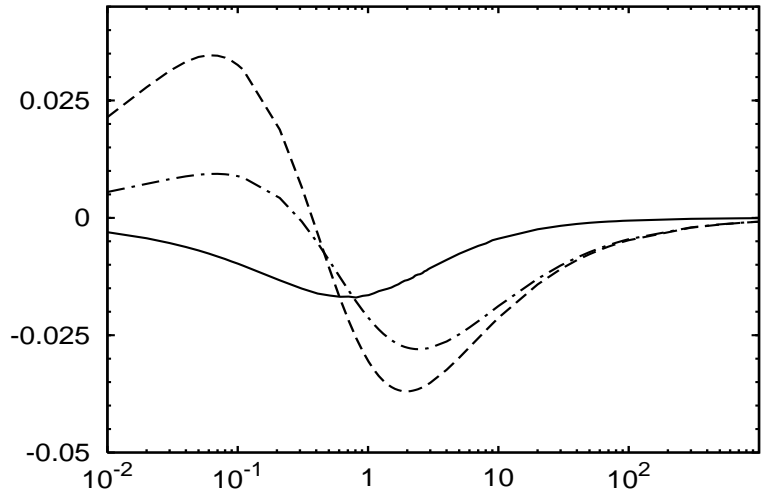, width=11.3cm, height=7cm}
\end{center}
\caption{Contributions of the initial and final vertex corrections
for $m_H$= 114 GeV (dashed), for $m_H$ = 250 GeV (dash-dotted), 
and of the box plus gluon radiation terms
(solid) to  $f^{(1)}_{d \bar d}(\eta)$.
}\label{fig:sigqcomp}
\end{figure}

In Fig. \ref{fig:sigqcomp} the contributions  to $f^{(1)}_{d\bar d}$ 
of the initial and final
vertex corrections and of the box  plus gluon radiation terms
are shown. These two contributions are separately infrared-finite. 
This figure clearly shows that the latter contributions should not be
neglected. This statement holds also for the spin observables
discussed below. We have numerically compared the contributions of the
initial and final vertex corrections to (\ref{eq:xsection}) relative
to the order $\alpha_s^2$ QCD Born cross section with the results Figs. 9 and
10 of \cite{Beenakker:1993yr} and find
agreement. Fig. \ref{fig:sigqcomp} also shows that for $\eta \lesssim
10$ the dependence on the Higgs boson mass is significant. 

The contribution of the  corrections shown in 
Fig. \ref{fig:sigqq}  to the $t \bar t$ cross section at the Tevatron is very small.
This is mainly due to the fact that the 
order $\alpha_s^2 \alpha$ corrections change sign for larger $\eta$.
The significance of the contributions can be enhanced by suitable
cuts, e.g. in the $t {\bar t}$ invariant mass.

\begin{figure}[H]
\begin{center}
\epsfig{file=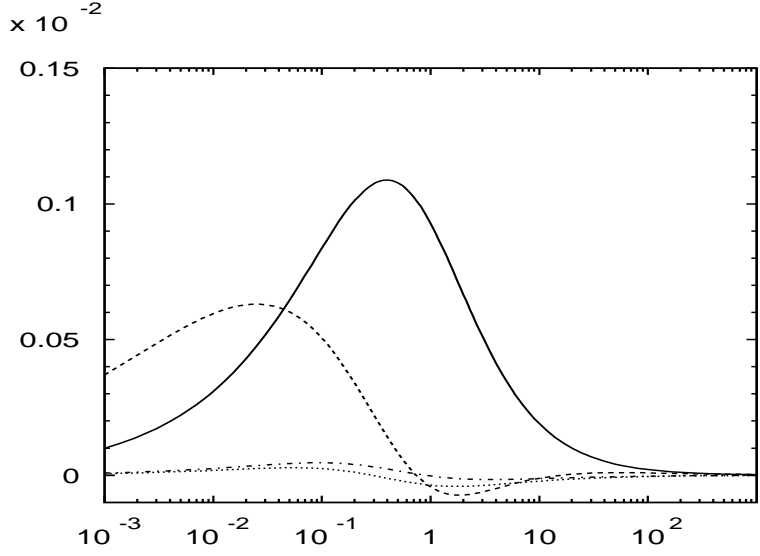, width=12cm, height=8cm}
\end{center}
\caption{ Contributions of the LO (solid) and NLO QCD (dashed)
contributions (taken
  from \cite{Bernreuther:2000yn}) and of the 
mixed $\alpha_s^2 \alpha$ contributions (dotted and dash-dotted
line refers to
initial d-type and u-type quarks, respectively) to the cross section
(\ref{mixed}) 
in units of $1/m_t^2$, and $m_H$ = 114 GeV.
}\label{Pcomparison}
\end{figure}

In Fig. \ref{Pcomparison}  the order $\alpha_s^2$, $\alpha_s^3$,  and
the $\alpha_s^2 \alpha$ contributions to the cross section
(\ref{mixed}) are shown as functions of $\eta$.  In these plots
$\alpha_s(m_t)=0.095$ and $\alpha(m_Z) = 0.008$ was chosen.
One sees that for $\eta \gtrsim 1$
the mixed corrections become of the same size or 
larger in magnitude than 
the NLO QCD contributions, and at $\eta \sim 10$ the
$\alpha_s^2 \alpha$ contributions are already about 15 percent of the 
LO QCD cross section.
These regions
can be investigated by studying the  distribution of the
$t {\bar t}$ invariant mass $M_{t \bar t}$. A value of, say,
 $\eta \sim 10$ 
corresponds roughly
to   $M_{t \bar t} \sim$ 1 TeV. For a quantitative discussion the 
reaction $gg\to t{\bar t}X$
must, of course,  also be taken into account.
 At the LHC, where such studies may be feasible,  this is the dominant
channel. \\ \\
Next we consider spin asymmetries. Denoting the top spin operator by
$\sp$ and its projection onto an arbitrary unit axis ${\bf\hat a}$
by $\sp \cdot {\bf\hat a}$ we can express 
its unnormalized  partonic expectation
value, which we denote by double brackets, in terms of the
difference between the ``spin up'' and ``spin down'' cross sections:
\begin{equation}
  2 \langle  \langle \sp \cdot {\bf\hat a}  \rangle  {\rangle}_i
  = \sigma_i(\uparrow ) - \sigma_i(\downarrow) \,  .
\label{sspas}
\end{equation}
Here $i = q \bar q$ and the arrows refer to the spin state of the top quark 
with  respect to ${\bf{\hat a}}$. An analogous formula  holds for the antitop
quark. It is these expressions  that enter the 
predictions for (anti)proton collisions.

There are two types
of single spin asymmetries (\ref{sspas}):
parity-even, T-odd asymmetries and
parity-violating, T-even  ones. The asymmetry associated with
the projection $\sp$ onto the normal of the $q, t$ scattering
plane belongs to the first class.
It is induced by the absorptive part of $\delta {\cal M}_2$,
but also by the absorptive part of the NLO QCD amplitude.  (This was
calculated in \cite{Bernreuther:1995cx,Dharmaratna:xd}.) The weak
contribution is even smaller  than the one from QCD which is of the
order of a few percent. For the sake of brevity we do not display it
here.  

\begin{figure}[H]
\begin{center}
\epsfig{file=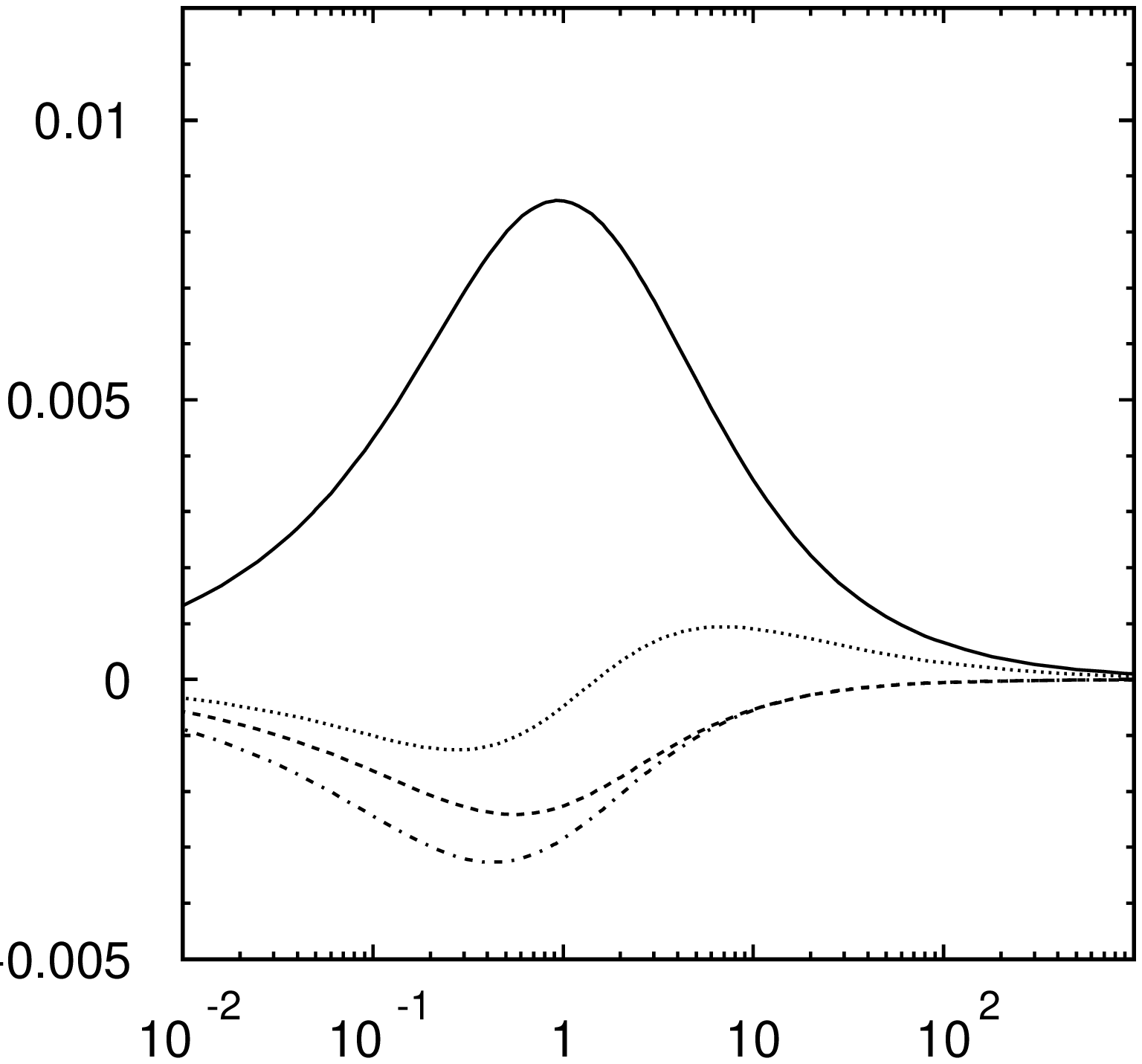, width=12cm,height=8cm }
\end{center}
\caption{Scaling functions $h^{(0,a)}_{q \bar q}(\eta)$
(dashed), $h^{(1,a)}_{q \bar q}(\eta)$ (solid)  that determine
the expectation value   (\ref{eq:sinspin}) for the beam axis in the
case
of $q=d$ type quarks. The dash-dotted and dotted lines correspond to
the respective functions for $q=u$ type quarks. 
$m_H$ = 114 GeV.}\label{fig:qqbeam}
\end{figure}

\begin{figure}[H]
\begin{center}
\epsfig{file=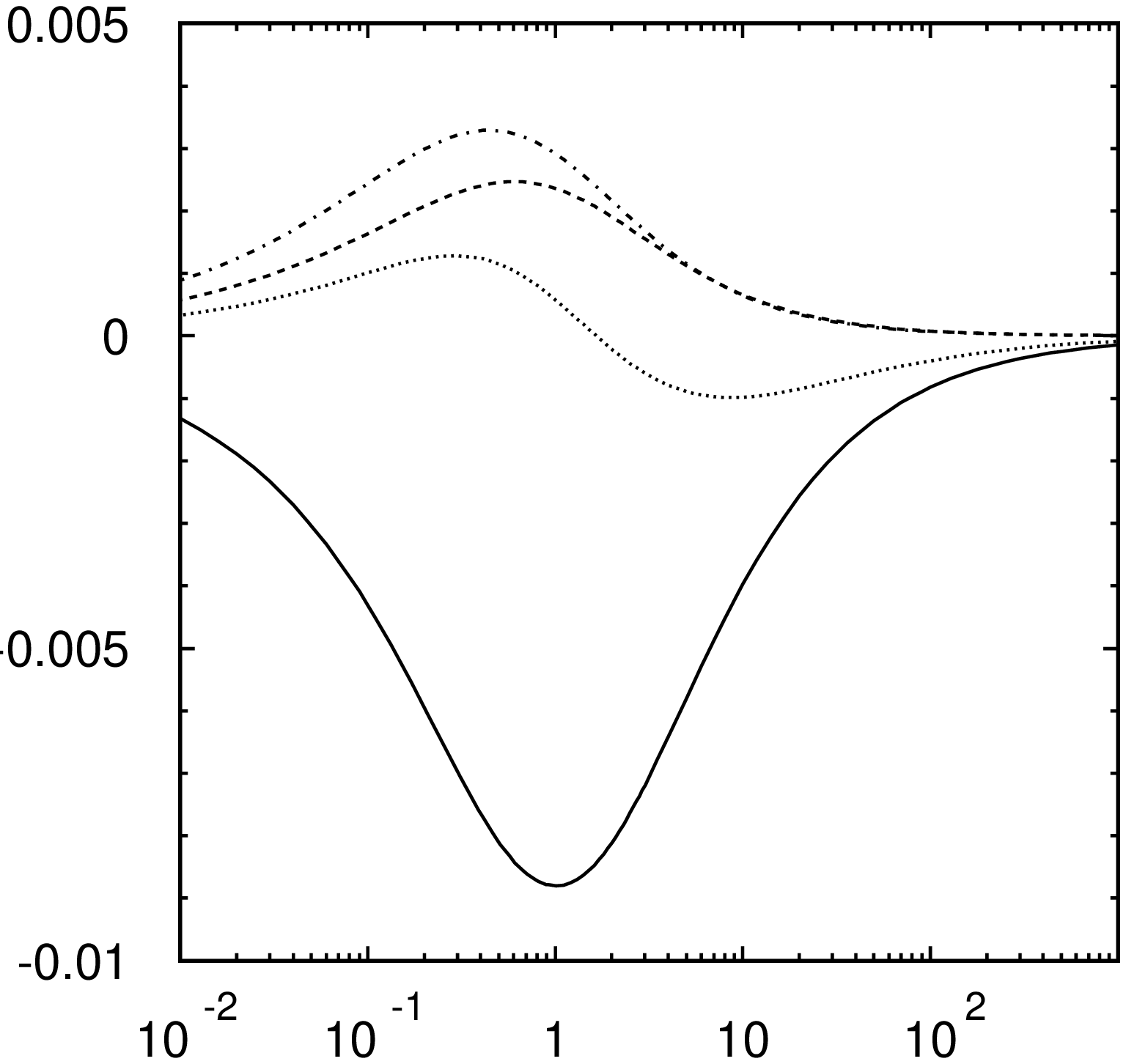, width=12cm,height=8cm }
\end{center}
\caption{Scaling functions $h^{(0,a)}_{q \bar q}(\eta)$
(dashed), $h^{(1,a)}_{q \bar q}(\eta)$ (solid)  that determine
the expectation value   (\ref{eq:sinspin}) for the off-diagonal  axis in the
case
of $q=d$ type quarks. The dash-dotted and dotted lines correspond to
the respective functions for $q=u$ type quarks. $m_H$ = 114 GeV.}\label{fig:qqoff}
\end{figure}

\begin{figure}[H]
\begin{center}
\epsfig{file=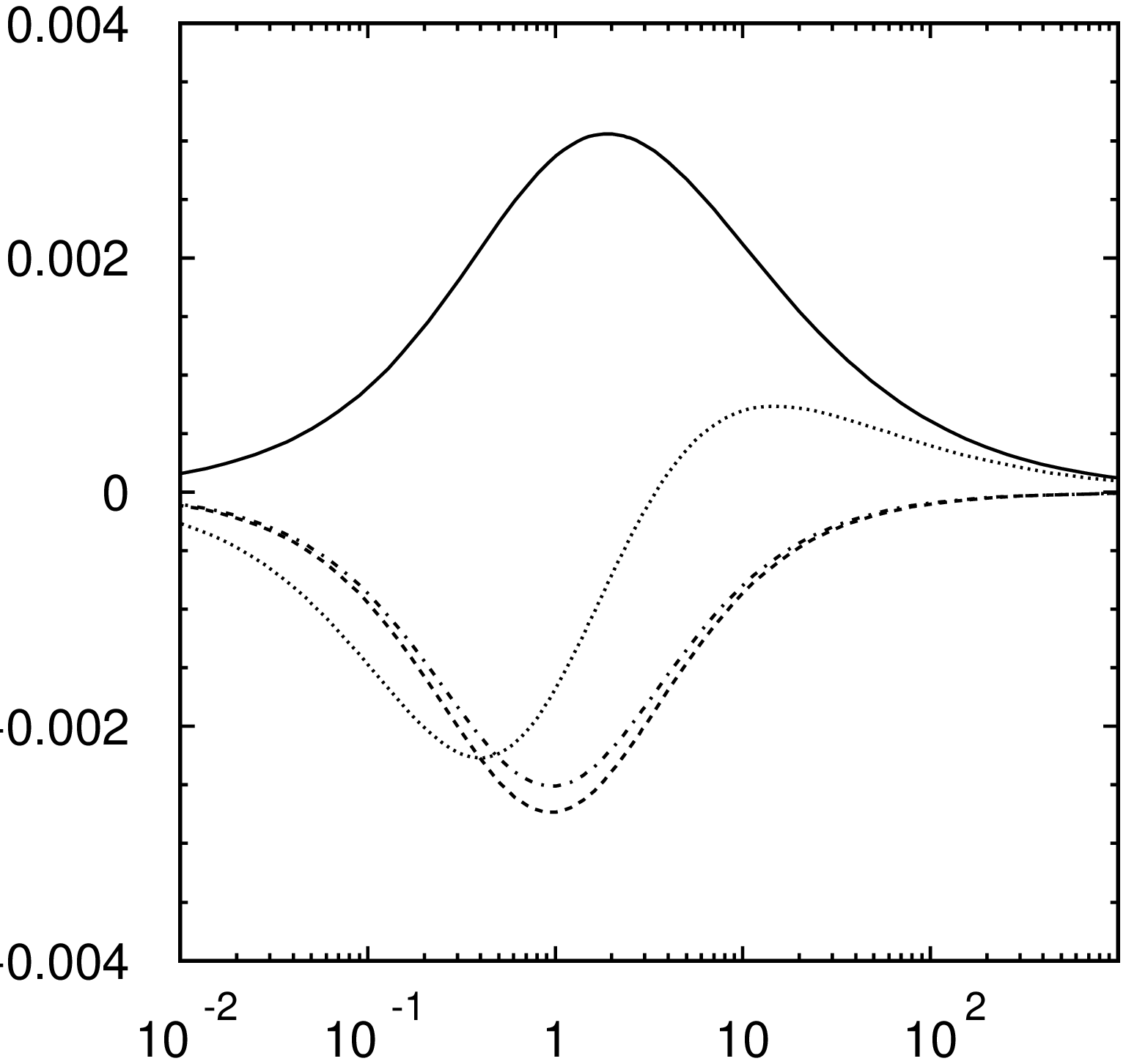, width=12cm,height=8cm }
\end{center}
\caption{Scaling functions $h^{(0,a)}_{q \bar q}(\eta)$
(dashed), $h^{(1,a)}_{q \bar q}(\eta)$ (solid)  that determine
the expectation value   (\ref{eq:sinspin}) for the helicity axis in the
case
of $q=d$ type quarks. The dash-dotted and dotted lines correspond to
the respective functions for $q=u$ type quarks. $m_H$ = 114GeV.}\label{fig:qqheli}
\end{figure}

The P-odd, T-even  single spin asymmetries correspond to projections of
the top spin onto a polar vector, in particular onto 
an axis that lies in the scattering plane. Needless to say, they
cannot be generated within QCD; the leading SM contributions to
these asymmetries are the parity-violating pieces of
eq. (\ref{convir}) and $\delta{\cal M}_3$ above. 
We consider top spin projections onto the beam axis (which is relevant
for the Tevatron), onto the helicity axis (of relevance for the LHC),
and for completeness also onto the so-called off-diagonal axis, which
was constructed to maximize $t \bar t$ spin correlations in  the
the  $q \bar q$ channel \cite{Mahlon:1997uc}. Naively, one might
define these  axes in the c.m. frame of the initial partons. However,
the observables  $\sp \cdot {\bf\hat a}$ are then not
collinear-safe. (The problem shows up once second-order QCD
corrections are taken into account.) A convenient frame with respect
to which collinear-safe spin projections can be defined is the $t \bar
t$ zero-momentum-frame (ZMF) \cite{Bernreuther:2004jv}.
With respect to this frame we define the axes  
\begin{eqnarray}
     {\bf\hat a} = {\bf\hat b} = {\bf\hat p}, &&\mbox{(beam\
      basis)}
\label{beambasis},\\
     {\bf\hat a} =  {\bf\hat b} = \dhh,&&
    \mbox{(off-diagonal\ basis)},
\label{offbasis} \\
     {\bf\hat a} = -  {\bf\hat b} = \kh,&&
    \mbox{(helicity axes)}
\label{helbasis}, 
\end{eqnarray}
where $\kh $ denotes the direction of
flight of the top quark in the  $t\bar{t}$-ZMF
and ${\bf\hat p}$ is the direction of flight of one of the colliding
hadrons in that frame. The direction of the hadron beam can be identified to
a very good approximation with the direction of flight of one of the
initial partons. The unit vectors ${\bf\hat b}$ are used for the
projections of the spin of the $\bar t$ quark.
The vector $\dhh$ is given by 
\begin{equation}
\dhh = \frac{-\ph+(1-\gamma)(\ph\cdot\kh)\kh}
{\sqrt{1-(\ph\cdot\kh)^2(1-\gamma^2)}} \, ,
\label{dopt}
\end{equation}
where  $\gamma=E/m$. \\
The unnormalized  expectation values of 
${\bf S_t}\cdot {\bf\hat a}$ are again conveniently expressed by scaling
functions
\begin{equation}
\langle \langle 2 {\sp}\cdot {\bf\hat a} \rangle
{\rangle}_{q \bar q}  =\frac{4\pi \alpha}{m^2_t}[
\alpha \, h^{(0,a)}_{q \bar q}(\eta) + \alpha_s^2 \, h^{(1,a)}_{q \bar q}(\eta)]
\, .
\label{eq:sinspin}
\end{equation}
The results for the scaling functions corresponding to the three axes
above are shown in Figs. \ref{fig:qqbeam} - \ref{fig:qqheli}.
For the beam and off-diagonal axes the asymmetries are  almost
equal both at LO and
at NLO, 
 up to sign, and for weak isospin $I_W =-1/2$ quarks the
$\alpha_s^2 \alpha$ corrections are significantly larger than the
LO values. This is in contrast to the helicity basis where the LO and
NLO terms shown in Fig. \ref{fig:qqheli} are of the same order of
magnitude. Moreover, in this basis the LO and NLO terms cancel each
other to a large extent for $I_W =-1/2$ quarks in the initial state.
Because $t \bar t$ production at the Tevatron occurs predominantly
by $q \bar q$ annihilation, these results imply that, when it comes to
searching for SM-induced parity-violating $t$ or $\bar t$ spin
effects, the reference axes of choice should be ${\bf\hat p}$ or $\dhh$. 
The SM effect is quite small; for suitably chosen $M_{t \bar t}$ mass
bins one gets asymmetries of the order of 2  percent. This leaves a
large margin in the search for new physics contributions. 

Finally  we consider top-antitop spin-spin correlations. For the sake
of brevity we concentrate here on 
parity- and T-even ones which are generated already to
lowest order QCD. For the Tevatron
these spin correlations (including NLO corrections) are
largest
with respect to the beam and off-diagonal bases, while for the
LHC the helicity basis is a good choice\footnote{For the LHC, a basis
has been  constructed \cite{Uwer:2004vp} which gives a QCD  effect which
is  somewhat larger than using  the helicity axes.}. In addition, a
good measure for the $t \bar t$ spin correlations at the LHC is, in
the case of the dilepton channels, the distribution
of  the opening angle between the charged leptons from semileptonic $t$
and $\bar t$ decay. At the level of $ t \bar t$ this amounts to the
correlation ${\bf S}_t\cdot {\bf S}_{\bar t}$ (for details, see 
 \cite{Bernreuther:2004jv}).
Therefore we consider the following set of
observables:
\begin{equation}
{\cal O}_1=4\,(\ph\cdot\sp)(\ph\cdot\sm),
\label{eq:bbasis}
\end{equation}
\begin{equation}
{\cal O}_2=4\,(\dhh\cdot\sp)(\dhh\cdot\sm),
\label{eq:obasis}
\end{equation}
\begin{equation}
{\cal O}_3=- 4\,(\kh\cdot\sp)(\kh \cdot\sm),
\label{eq:hbasis}
\end{equation}
\begin{equation}
{\cal O}_4=4\,\sp\cdot\sm,
\label{eq:sbasis}
\end{equation}
where the axes are as defined  in eqs.  (\ref{beambasis})
-  (\ref{helbasis})  in the $t \bar t$ ZMF
and the factor 4 is conventional.
The unnormalized expectation
values of these observables correspond to unnormalized double spin
asymmetries, i.e.,
to the following combination of $t, \bar t$ 
spin-dependent  cross sections: 
\begin{equation}
\label{double}
  \langle \langle {\cal O}_b\rangle
 \rangle_i
 \;  = \; \sigma_i (\uparrow \uparrow)+\sigma_i(\downarrow \downarrow)
  - \sigma_i(\uparrow \downarrow)- \sigma_i(\downarrow 
\uparrow) \, ,
\label{dcorrw}
\end{equation}
and here $i={q \bar q}$.
The arrows on the right-hand side refer to the spin state of the top 
and antitop quarks  with respect to the  reference 
axes $ {\bf \hat a}$ and $ {\bf \hat b}$. 

\begin{figure}[H]
\begin{center}
\epsfig{file=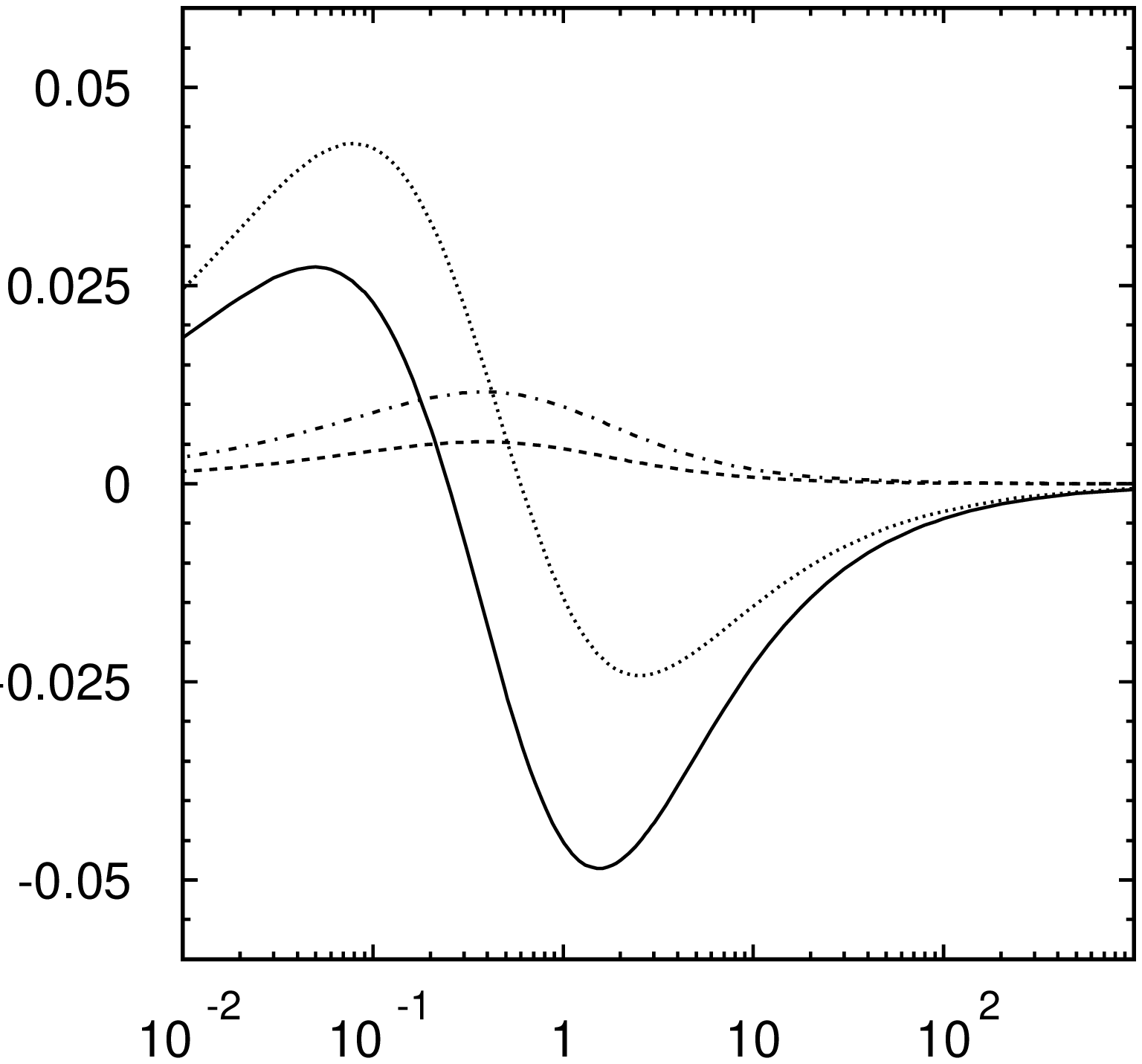, width=12cm,height=8cm }
\end{center}
\caption{Scaling functions $g^{(0,1)}_{q \bar q}(\eta)$
(dashed), $g^{(1,1)}_{q \bar q}(\eta)$ (solid)  that determine
the expectation value   (\ref{eq:douspin}) for the beam basis in the
case
of $q=d$ type quarks. The dash-dotted and dotted lines correspond to
the respective functions for $q=u$ type quarks.  $m_H$ = 114 GeV.}\label{fig:qdbeam}
\end{figure}

\begin{figure}[H]
\begin{center}
\epsfig{file=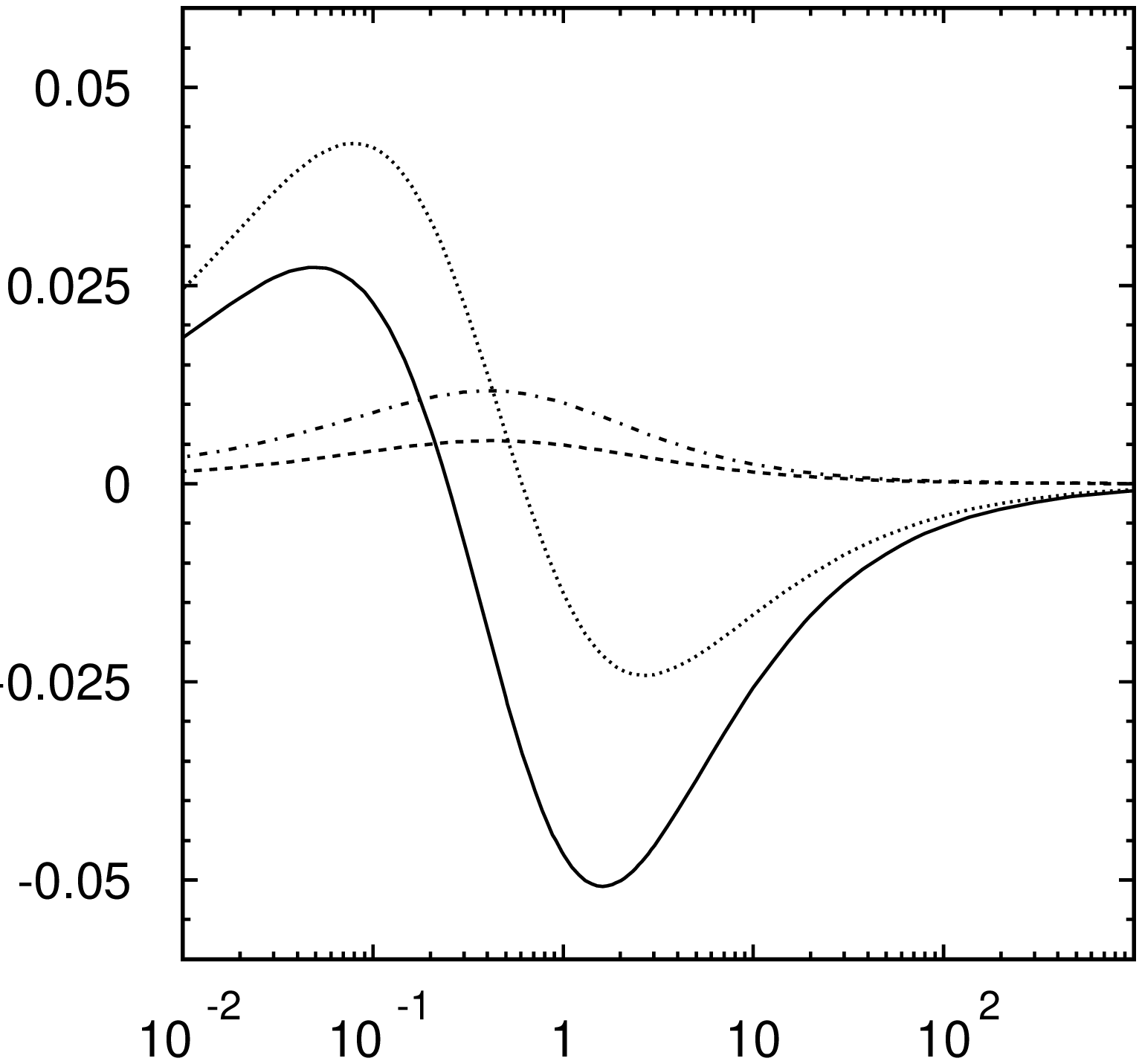, width=12cm,height=8cm }
\end{center}
\caption{Scaling functions $g^{(0,2)}_{q \bar q}(\eta)$
(dashed), $g^{(1,2)}_{q \bar q}(\eta)$ (solid)  that determine
the expectation value   (\ref{eq:douspin}) for the off-diagonal  basis in the
case
of $q=d$ type quarks. The dash-dotted and dotted lines correspond to
the respective functions for $q=u$ type quarks.  $m_H$ = 114 GeV.}\label{fig:qdoff}
\end{figure}

\begin{figure}[H]
\begin{center}
\epsfig{file=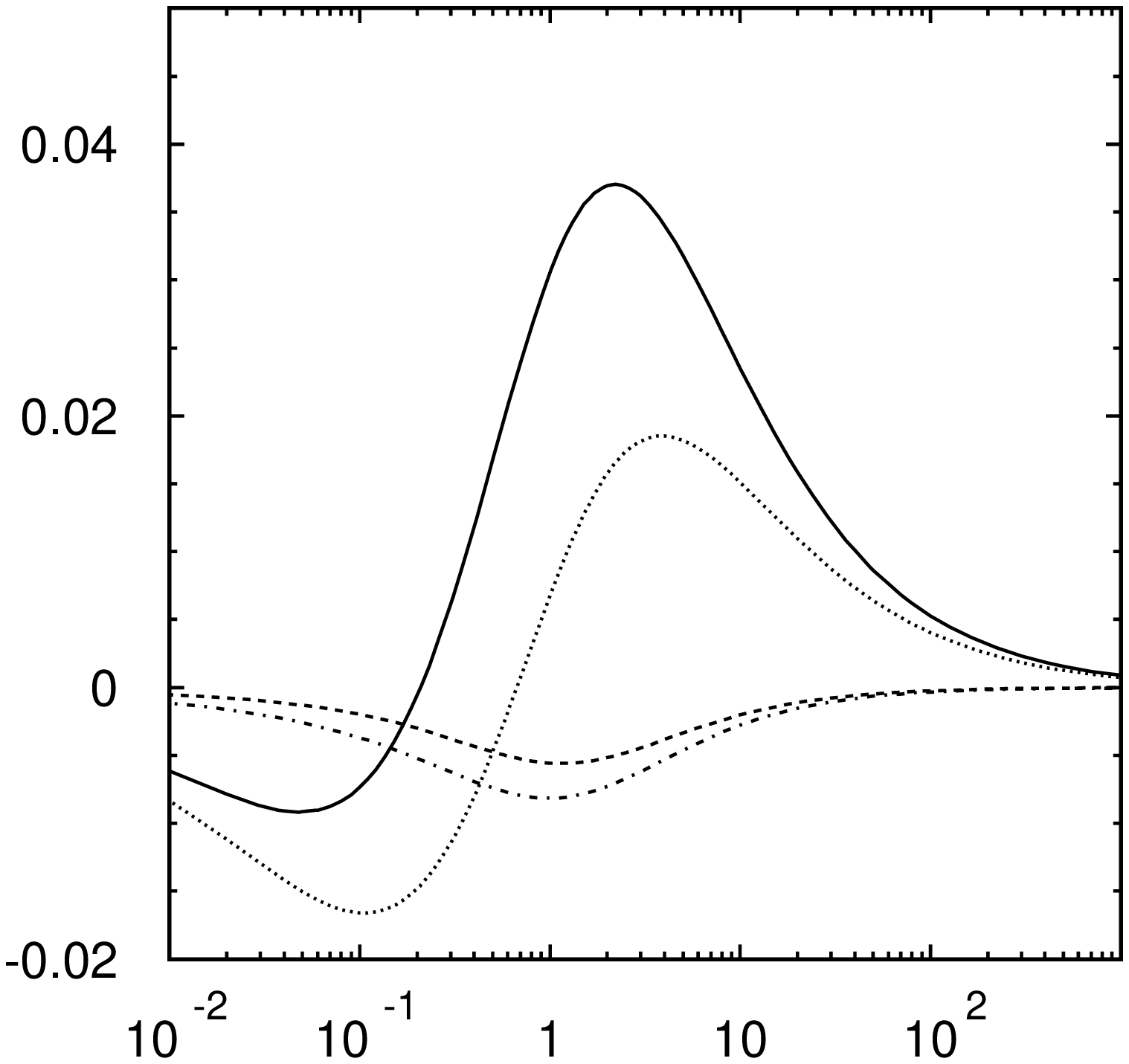, width=12cm,height=8cm }
\end{center}
\caption{Scaling functions $g^{(0,3)}_{q \bar q}(\eta)$
(dashed), $g^{(1,3)}_{q \bar q}(\eta)$ (solid)  that determine
the expectation value   (\ref{eq:douspin}) for the helicity basis in the
case of $q=d$ type quarks. The dash-dotted and dotted lines correspond to
the respective functions for $q=u$ type quarks.  $m_H$ = 114 GeV.}\label{fig:qdheli}
 \end{figure}

Again we compute the $\alpha^2$ and weak-QCD contributions of
order  $\alpha_s^2 \alpha$ to (\ref{dcorrw})
and express them in terms of scaling functions: 
\begin{equation}
\langle \langle {\cal O}_b
\rangle {\rangle}_{q \bar q}^W  =\frac{4\pi \alpha}{m^2_t}[
\alpha \, g^{(0,b)}_{q \bar q}(\eta) + \alpha_s^2 \, g^{(1,b)}_{q \bar q}(\eta)]
\, .
\label{eq:douspin}
\end{equation} 
These functions are plotted in Figures \ref{fig:qdbeam}  -
\ref{fig:qdheli}, respectively. As is the case in QCD, in $ q \bar q$
annihilation the
spin correlations in the beam and  off-diagonal bases, and in the  projection 
(\ref{eq:sbasis}) are not very much different from each other.
 The size of the mixed corrections
(\ref{eq:douspin}) is  typically only a few percent  compared with the
QCD contributions  \cite{Bernreuther:2004jv} to (\ref{dcorrw}).
\par
The unnormalized expectation value of ${\cal O}_4$ given in
eq. (\ref{eq:sbasis}) is  equal to the respective contribution 
$\delta {\sigma}_{q \bar q}^W$ displayed in Fig. \ref{fig:sigqq}.
The reason is as follows. The S matrix elements for the reactions
$q {\bar q} \to t {\bar t} (g)$, to the order in the couplings
considered above, contain all possible partial wave amplitudes.
However, in the expectation value of the parity-even operator
(\ref{eq:sbasis})
only the $^3S_1$ (in the Born terms and the terms of
order  $\alpha_s^2 \alpha$) and $^3P_1$ components (in the Born term from $Z$
boson exchange)
 of the $t \bar t$ state contribute, as a closer inspection
shows. It is then a simple
exercise to show that the normalized 
expectation value of ${\cal O}_4$ is equal to one in this case.
Thus, its unnormalized expectation value is equal  to $\delta
{\sigma}_{q \bar q}^W$, which is confirmed by explicit calculation. (See
\cite{Bernreuther:1997gs} for similar considerations.)
\par
Another interesting class of asymmetries are parity-violating
double spin asymmetries of the form \\
$\delta A({\bf \hat a},{\bf \hat b}) =   
\sigma_i(\uparrow \downarrow)- \sigma_i(\downarrow 
\uparrow)$.  They  are generated by the  parity-violating pieces of
(\ref{convir}) and of $\delta{\cal M}_3$ above. In \cite{Kao:1999kj}
an observable of this form was computed in the $t, \bar t$ helicity
basis within the SM, with box plus gluon contributions not taken into
account, and in some SM extensions. 

In addition, the absorptive parts of $\delta {\cal M}_2$ lead to T-odd
spin-spin correlations, both P-even and odd ones. These are, however,
very small effects, and we do not display them here. 

The single and double spin asymmetries are reflected in respective
angular distributions/asymmetries  of the $t$ and $\bar t$ decay
products.
In particular they  contribute to the one- and two-particle inclusive
decay distributions $\sigma^{-1} d\,\sigma/d\,\cos\theta_1$ and 
 $\sigma^{-1} d\,\sigma/(d\,\cos\theta_1\, d\,\cos\theta_2)$,
where  $\theta_1, \theta_2$ are the angles between the
 direction of flight of a $t$ and $\bar t$ decay product,
 respectively, and a chosen
reference  direction. Suitable reference directions are 
the axes introduced above. The effects are largest if the charged
lepton(s) from $t$ and/or $\bar t$ decay is (are) used as spin
analyzer(s). The
weak contributions can be enhanced with respect to the pure QCD
effects by  suitable cuts in the $t\bar t$ invariant mass. 
Numerical studies, including the weak contributions to
$gg \to t \bar t$ will be given elsewhere \cite{BFS05}.

In summary we have computed the mixed QCD and weak corrections to top
quark pair production by  quark antiquark annihilation, keeping the
full dependence on the $t$ and $\bar t$ spins. These results,
combined with 
our previous QCD results and with the mixed contributions to $ gg \to
t \bar t$,
will allow for detailed predictions of top quark observables, in
particular of
top spin-induced angular correlations and asymmetries within the
standard model \cite{BFS05}.
Specifically, the results of this letter are necessary ingredients 
for SM predictions of  parity-violating observables associated with the
spin of the (anti)top quark. 
We believe that such observables, which are very sensitive
to non-SM parity-violating top quark interactions, 
will become  important analysis tools  once sufficiently large $t
 \bar t$ data samples will have been collected.

\subsubsection*{Acknowledgements}
We wish to thank Arnd Brandenburg for helpful discussions,
and A. Scharf and P. Uwer for discussions and 
for communication of their results
prior to publication.  This work was
supported
by Deutsche Forschungsgemeinschaft (DFG) SFB/TR9, by
DFG-Graduiertenkolleg RWTH Aachen, and by
the National Natural Science Foundation of China (NSFC). Z.G. Si
thanks NSFC and DFG  for financial support during his stay at RWTH
Aachen.

\newpage
\newpage
\vspace*{-1cm}
\newpage
\newpage
 \newpage

\end{document}